\documentclass{amsart}
\usepackage{amsmath, amsthm, amsfonts}
\usepackage{mathrsfs}
\usepackage[colorlinks=false,linktocpage=true]{hyperref}
\usepackage{tocvsec2}
\usepackage{latexsym}
\font\tenscrpt=eusm10
\font\sevenscrpt=eusm10 scaled 700
\font\fivescrpt=eusm10 scaled 500
\newfam\eusmfam
\textfont\eusmfam=\tenscrpt
\scriptfont\eusmfam=\sevenscrpt
\scriptscriptfont\eusmfam=\fivescrpt
\def\scr#1{{\fam\eusmfam\relax#1}}
\newtheorem{thm}{Theorem}[section]
\newtheorem{cor}{Corollary}[section]
\newtheorem{lem}{Lemma}[section]

\newtheorem{rem}{Remark}[section]

\newcommand{\thmref}[1]{Theorem~{\rm $\ref{#1}$}}

\newcommand{\lemref}[1]{Lemma~{\rm $\ref{#1}$}}

\newcommand{\remref}[1]{Remark~{\rm $\ref{#1}$}}

\numberwithin{equation}{section}
\def\eqdef{\overset{\triangle}{=}}
\def\P{{\mathbb P}}
\def\Pt{\tilde{\mathbb P}}
\def\EP{{\mathbb E}_{\P}}

\def\Wm{\scr W}
\def\sB{\scr B}
\def\sF{\scr F}
\def\sFt{{\scr F}_t}
\begin{document}
\title[Space-Time Change of Measure Look at Market Price via semi-SPDE's]{Market Price of Risk and Random Field Driven
Models of Term Structure: A Space-Time Change of Measure Look}
\author{Hassan Allouba}
\address{Department of Mathematics, Indiana University, Bloomington, IN 47405}
\curraddr{Dept. of Mathematical Sciences, Kent State University, Kent, OH 44240}
\email{allouba@indiana.edu}
\author{Victor Goodman}
\address{Department of Mathematics, Indiana University, Bloomington, IN 47405}
\email{goodmanv@indiana.edu}
\date{October 4, 2001}
 \begin{abstract}
No-arbitrage models of term structure have the feature that the
return on zero-coupon bonds is the sum of the short rate and the product of
volatility and market price of risk.  Well known models restrict the behavior
of the market price of risk so that it is not dependent on the type of asset being
modeled.  We show that the models recently proposed by Goldstein and Santa-Clara and Sornette, among others,
allow the market price of risk to depend on characteristics of each asset, and we
quantify this dependence.    A key tool in our analysis is a very general space-time change of measure theorem, proved by the first author in earlier work,
and covers continuous orthogonal local martingale measures including space-time white noise.
\end{abstract}
\maketitle

\section{Introduction}
Let $P(t,T)$ denote the market price of a discount bond that matures at time $T$.
Since bond trading prices share many of the same characteristics of stock prices ,
several approaches to modeling bond prices use a stochastic noise term to express
the uncertainty regarding future prices of a specific bond. The approach
in Hull and White (1990), is to model each discount bond with an SDE of the form
\begin{equation}
dP = \mu(t,T)Pdt + P\sigma(t,T) dW
\label{SDE}
\end{equation}
where $W(t)$ is a one-parameter Brownian motion which serves as a
shared noise term.  On the other hand, the coefficients $\mu$ and $\sigma$ are
adapted functions of $t$ which also depend on the maturity time $T$.  That is,
these terms capture characteristics that are unique to the different maturity
dates.
By making some technical assumptions regarding $\mu$ and $\sigma$, one can derive a
short--term interest rate process, $r(t)$ from the bond prices as well as forward
interest rates, which form the modeling equations of Heath, Jarrow, and Morton
term structure models \cite{HJM}.  HJM interest rate models are consistent with such SDE
families given by \eqref{SDE} (see Baxter, Rennie \cite{BR}).

One attempts to choose the``parameters'' $\mu$ and $\sigma$ so that various
correlations of bond price behavior can be attained while a consistency of bond
prices is maintained.  By this, we mean that arbitrage opportunities are ruled out
within the model.
\subsection{Condition for no--arbitrage} The drift $\mu(t,T)$, for each bond
maturity is a function of the short (interest) rate and the {\it market price of
risk}, a single function $\lambda(t)$ that determines the underlying
return on each bond price:
There is an adapted function $\lambda(t,\omega)$, such that for $0\leq t\leq T$,
$$\mu(t,T) = r(t) + \lambda(t,\omega)\sigma(t,T)$$
This peculiar feature of one-factor as well as multi-factor bond models
states that the ``excess'' return on bonds of different maturities are
proportional to their volatilities.  This seems an unrealistic oversimplification that fails to take into account
different characteristics of bonds with different maturities.

More elaborate bond models have been proposed by Goldstein, Santa-Clara and Sornette, and others.  These
models replace a Brownian motion by a Gaussian random field, $Z(t,T)$.  More
general correlations of bond prices may be modeled with these random field
noises.

We show that a more general market price of risk is compatible with no-arbitrage
models for a variety of random fields generated by the two-parameter Brownian
sheet process.  We also characterize the generality of market price of risk within
these classes of models.

\section{Random Field Driven Bond Models}
Assume that the random field $Z(t,T)$ is a Brownian motion for each fixed $T$, and that the process
is continuous in both variables.  We use the bond model
\begin{equation}
dP = \mu(t,T)Pdt + P\sigma(t,T) d_tZ(t,T)
\label{HalfSPDE}
\end{equation}
where all differentials, including $d_tZ(t,T)$, are taken in the $t$ variable only; and so we will often refer to \eqref{HalfSPDE} as a semi-SPDE or a parametrized SDE.
We assume throughout the rest of the article that $\sigma$ is continuous in $t$.  In order to find sufficient conditions (on a market price of risk) so that no arbitrage is possible, we express the drift in terms of an existing short rate
process $r(t)$ and an unknown function $\lambda$ as follows:
For some adapted function $\lambda(t,T,\omega)$, of both time and maturity date, we have
$$\mu(t,T) = r(t) + \lambda(t,T,\omega)\sigma(t,T)$$
We find that {\it some dependence} on $T$ is consistent with no arbitrage, and we obtain the space-time risk neutral measure.
\subsection{Main Result}
The following Theorem illustrates the generality of allowable market prices of risk
for certain random field term structure models.     It addresses the bond model in \eqref{HalfSPDE} when the noise $Z$ has the form
\begin{equation}
Z\eqdef W(t,T)/{\sqrt{T}}=\Wm_t([0,T])/{\sqrt{T}}
\label{orn}
\end{equation}
where $W(t,T)$ is the two-parameter Brownian sheet (a zero mean Gaussian process with $Cov(W(t_1,T_1),W(t_2,T_2))=(t_1\wedge t_2)(T_1\wedge T_2)$) corresponding to a space-time white noise
$\Wm\eqdef\{\Wm_t(B),\sFt;0\le t\le T_0, B\in\sB([0,T_0])\}$ on a usual probability space $(\Omega,\sF,\{\sFt\},\P)$ ($\sB([0,T_0])$ is the Borel $\sigma$-field over $[0,T_0]$).
The relation between space-time white noise as a local martingale measure and its induced Brownian sheet and the relevant definitions are detailed in \cite{A,W},
and we refer the interested reader to these references for additional interesting details.
\begin{thm}
Suppose that $Z$ has the form in \eqref{orn}.  Then for any $T_0 > 0$ and any market price of risk $\lambda$ of the form
\begin{equation}
\lambda(t,T)=\int_0^T\eta(t,u, \omega )du;\hskip.4in\  T \leq T_0,
\label{PR}
\end{equation}
where $\eta$ is an $\sFt$-predictable random field $($see \cite{A,W}\/$)$ such that
\begin{equation}
\EP\exp\left(\int_0^{T_0}\int_0^{T_0}\big[\log(T/u)\eta(t,u)\lambda(t,u)/2+u\eta(t,u)^2\big]\  dudt \right)<\infty,
\label{C1}
\end{equation}
the bond model \eqref{HalfSPDE} is free from arbitrage over the time interval $0\leq t\leq T_0$.
\label{Arb1}
\end{thm}
\begin{rem} We make the following observations:
\begin{enumerate}
\item This result shows that no-arbitrage is consistent with
market prices of risk that are $T$-dependent and also that the market price of risk
is absolutely continuous with respect to the maturity time.
\item As shown in the next subsection, as a Corollary of the proof of \thmref{Arb1}, the integrability condition in \eqref{C1} may be replaced by the
simpler condition
\begin{equation}
\EP\exp\left({5T_0\over 4}\int_0^{T_0}\int_0^{T_0}\eta(t,u)^2\  dudt\right)<\infty
\label{C2}
\end{equation}
\item The space-time setting we are adopting here and in \lemref{STCOM} is more natural and flexible than the classical multiparameter one: it gives time its traditional role in processes while
allowing the space variable $($in this case the maturity date\/$)$ to be free to take from any space, not necessarily symmetric to the time set.   It can also accomodate
a larger class of noises while avoiding the unnatural restrictions imposed by the multiparameter setting $($see \cite{A}\/$)$.   This makes our results extendable to more general models.
\item The covariance structure of the random field $Z(t,T)$ in \thmref{Arb1} is, of course,
$$\EP[Z(t_1,T_1)Z(t_2,T_2)]= (t_1\wedge  t_2)\sqrt{{(T_1\wedge  T_2)\over(T_1\vee  T_2)}}$$
\end{enumerate}
\label{remarks}
\end{rem}
\subsection{Proofs and Extensions of the Main Result}
We start with the aforementioned space-time change of measure result needed in our proof of \thmref{Arb1}.
This is a special case of Corollary 2.3 along with Lemma 2.4 in \cite{A}, which we combine and state here for the reader's convenience as
\begin{lem}
Suppose that $\Wm=\{\Wm_t(B),\sFt;0\le t\le T_0, B\in\sB([0,T_0])\}$ is a space-time white noise on the usual probability space $(\Omega,\sF,\{\sFt\},\P)$.
Suppose further that  $g$ is an $\sFt$-predictable random field satisfying
\begin{equation}
\EP\exp\left({1\over 2}\int_0^{T_0}\int_0^{T_0}\ g(s,u)^2\  dsdu\right)<\infty,
\label{N}
\end{equation}
then
$$\tilde{\Wm}_t(B)\eqdef \Wm_t(B) + \int_0^t \int_B g(s,u)duds;\qquad 0\le t\le T_0, B\in\sB([0,T_0])$$
is a white noise on $(\Omega,\sF_{T_0},\{\sFt\},\Pt),$ where
$${d\Pt\over d\P}= \exp\left[-\int_0^{T_0}\int_0^{T_0}g(s,u)\Wm(ds,du) -1/2\int_0^{T_0}\int_0^{T_0}g(s,u)^2dsdu\right].$$
I.e., the random field $\left\{\tilde{W}(t,T)\eqdef\tilde{\Wm}_t([0,T]);0\le t,T\le T_0\right\}$ is a Brownian sheet with respect to $\Pt$.
\label{STCOM}
\end{lem}
We now turn to the
\begin{proof}[Proof of \thmref{Arb1}]
We assume that the market price of risk $\lambda$ satisfies \eqref{PR}, and we
consider the process
\begin{equation}
\begin{split}
\tilde{Z}(t,T)\eqdef Z(t,T) + \int_0^t\lambda(s,T)ds={1\over \sqrt{T}}W(t,T) + \int_0^t\int_0^T\eta(s,u)duds
\end{split}
\label{newsheet}
\end{equation}

We wish to determine an absolutely continuous probability measure $\Pt$ defined on the $\sigma$-field $\sF_{T_0}$ so that the process in
\eqref{newsheet} is a martingale for each fixed $T>0$.  For this purpose, it suffices to find $\Pt$  making the process
\begin{equation}
\tilde{W}(t,T)\eqdef W(t,T) + \sqrt{T}\int_0^t \int_0^T\eta(s,u)duds
\label{newsheet2}
\end{equation}
a Brownian sheet over the parameter range $[0,T_0]\times[0,T_0]$ (each process in \eqref{newsheet}, for fixed $T$, is then a standard Brownian
motion with respect to such a measure $\Pt$).   Towards this end, we need to choose $g$ satisfying the generalized Novikov condition \eqref{N} such that
\begin{equation}
 \int_0^t \int_0^T g(s,u)duds=\sqrt{T}\int_0^t \int_0^T\eta(s,u)duds
\label{drift}
\end{equation}
Letting
\begin{equation}
g(s,u)\eqdef\frac{d}{du}\left(\sqrt{u}\int_0^u\eta(s,y)dy\right)={1\over 2\sqrt{u}}\lambda(s,u)+ \sqrt{u}\eta(s,u),
\label{gchoice}
\end{equation}
where $\lambda(s,u)=\int_0^u\eta(s,y)dy$, it is formally clear that \eqref{drift} holds.   We verify the validity of this formal computation by
computing the $L^2[0,T_0]$ norm of each term in the function $g$, and we will show that each $L^2$ norm is finite a.s.
In fact, we will compute the $L^2$ norm in both variables $t$ and $T$ over the square.

First,
\begin{equation*}
\begin{split}
\left\|{1\over 2\sqrt{u}}\lambda(s,u)\right\|^2_2&= \int_0^{T_0}\int_0^{T_0}{1\over 4u}\lambda^2(s,u)duds\\
&=\int_0^{T_0}\int_0^{T_0}{1\over 4u}\int_0^u\int_0^u\eta(s,r)\eta(s,\tau)drd\tau duds\\
&=\int_0^{T_0}\int_0^{T_0}\int_0^{T_0}\eta(s,r)\eta(s,\tau)\int_{r\vee \tau}^{T_0}{1\over 4u}du drd\tau ds\\
&={1\over 4}\int_0^{T_0}\int_0^{T_0}\int_0^{T_0}\eta(s,r)\eta(s,\tau)\log\left({T_0\over {r\vee \tau}}\right)drd\tau ds\\
&={1\over 2}\int_0^{T_0}\int_0^{T_0}\eta(s,\tau)\int_0^{\tau}\eta(s,r)\log\left({T_0\over {\tau}}\right)drd\tau ds\\
&={1\over 2}\int_0^{T_0}\int_0^{T_0}\eta(s,\tau)\lambda(s,\tau)\log\left({T_0\over {\tau}}\right)d\tau ds
\end{split}
\end{equation*}
This quantity is finite a.s., since \eqref{C1} implies that its exponential has
finite mean.  Secondly,
$$\left\|\sqrt{u}\eta(s,u)\right\|^2_2 =\int_0^{T_0}\int_0^{T_0}u\eta(s,u)^2dsdu$$
is also finite a.s.~for the same reason.

Now the condition \eqref{N} of \lemref{STCOM} may be stated as $\EP\exp(\|g\|^2_2/2)<\infty$.  We use the inequality
$$\|g\|^2_2=\left\|{1\over 2\sqrt{u}}\lambda(s,u) + \sqrt{u}\eta(s,u)\right\|^2_2\leq 2\left\|{1\over 2\sqrt{u}}\lambda(s,u)\right\|^2_2+
2\left\| \sqrt{u}\eta(s,u)\right\|^2_2$$
and the estimates previously derived to obtain
$$\frac{\left\|g\right\|^2_2}{2}\leq {1\over 2}\int_0^{T_0}\int_0^{T_0}\eta(s,u)
\lambda(s,u)\log\left({T_0\over {u}}\right)du ds+\int_0^{T_0}\int_0^{T_0}u\eta(s,u)^2dsdu$$
But, the assumption in \eqref{C1} implies that the exponential moment of this quantity is finite;
hence, the hypothesis of  \lemref{STCOM} is satisfied.

Consider a bond model where the dynamics of discount bonds are given by \eqref{HalfSPDE} and where $\mu$ has the form $\mu = r +\lambda \sigma$.  For each $T>0$ we have
\begin{equation}
dP = (r(t)+ \lambda(t,T)\sigma(t,T))Pdt + \sigma(t,T)P d_tZ(t,T)\label{SSPDE2}
\end{equation}
where we assume that $(P,Z)$ is a solution of this semi-SPDE (parametrized SDE) on $(\Omega,\sF,\{\sFt\},\P)$.  We let $\tilde Z(t,T)$
denote the process defined in \eqref{newsheet}, and we note that
$$Z(t,T)= \tilde Z(t,T) - \int_0^t\lambda(s,T)ds$$
Under the measure $\Pt$ of  \lemref{STCOM},  $\sqrt{T}\tilde Z(t,T)$ is a Brownian sheet, so that  $\tilde Z(t,T)$
is a standard Brownian motion for fixed $T$.  Then the integral form of  \eqref{SSPDE2}, which holds pathwise, may be written in differential form as
\begin{equation}
\begin{split}
dP &= (r(t)+ \lambda(t,T)\sigma(t,T))Pdt + \sigma(t,T)P d_t\tilde Z(t,T)- \lambda(t,T)\sigma(t,T)Pdt\\
&= r(t)Pdt+ \sigma(t,T)P d_t\tilde Z(t,T)
\label{SSPDE3}
\end{split}
\end{equation}
We now proceed to verify Harrison and Kreps criteria for no arbitrage \cite{HaK}.  Since the semi-SPDE defined by \eqref{SSPDE3} is in terms of a standard Brownian motion $\tilde{Z}$,
one may fix $T$ and use the It\^o formula to investigate each process
$$D_t\eqdef\exp\left[-\int_0^t r(s)ds\right]P(t,T)$$
It follows that
$$dD_t=\exp\left[-\int_0^t r(s)ds\right]\sigma(t,T)P(t,T)d_t\tilde Z(t,T)$$
and hence $D_t$ is a martingale in $t$, under $\Pt$ (since $\sigma$ is assumed continuous in $t$, and thus bounded on $[0,T_0]$). The model then satisfies the desired  Harrison and Kreps criteria developed in \cite{HaK} to form a model
with no arbitrage and \thmref{Arb1} is proved.
\end{proof}
We now prove the second observation in \remref{remarks}
\begin{cor}
  In \thmref{Arb1}, if we replace the integrability condition \eqref{C1} by
  \begin{equation}
\EP\exp\left({5T_0\over 4}\int_0^{T_0}\int_0^{T_0}\eta(t,u)^2\  dudt \right)<\infty
\label{sC1}
\end{equation}
then \thmref{Arb1} holds.
\label{obs2}
\end{cor}
\bigskip
\begin{proof} We see from \eqref{drift} and \eqref{gchoice} and the following discussion that if
$$g(t,T)\eqdef{1\over 2\sqrt{T}}\int_0^T\eta(t,u)du + \sqrt{T}\eta(t,T)$$
it suffices to show
$$\EP\exp\left[\ {1\over 2}\|g\|^2_2\right]<\infty.$$
Applying the Cauchy-Schwarz inequality, we obtain
$$\left|{1\over 2\sqrt{T}}\int_0^T\eta(t,u)du\right|^2
\leq {1\over 4T}\int_0^T\eta^2(t,u)du\cdot T\leq{1\over 4}\int_0^{T_0}\eta^2(t,u)du$$
Then
$$\left\|{1\over 2\sqrt{T}}\int_0^T\eta(t,u)du\right\|^2_2\leq {T_0\over 4}\|\eta\|^2_2$$
so that
$$\|g\|^2_2\leq 2{T_0\over 4}\|\eta\|^2_2+ 2T_0\|\eta\|^2_2=  {5T_0\over 2}\|\eta\|^2_2$$
Now,
$$\EP\exp\left[ {1\over 2}\|g\|^2_2\right]\le\EP\exp\left[{5T_0\over 4}\|\eta\|^2_2\right], $$
and that last term is finite by assumption \eqref{sC1}.
\end{proof}

In Santa-Clara and Sornette \cite{SCSo}, more general random field term structure models are considered.  These use a noise term of the form
\begin{equation}
\label{newn}
Z(t,T)= {1\over h(T)}W(t,h^2(T))={1\over h(T)}\Wm_t([0,h^2(T)])
\end{equation}
where $h$ is an increasing, positive function on the interval $[0,T_0]$.    Clearly, such random fields have the feature that for each fixed $T$, the process is a
standard Brownian motion.  Also, the random field correlation is more general than for the normalized Brownian sheet in \eqref{orn}.  \thmref{Arb2} below
shows that such models allow essentially the same type of market price of risk behavior as in \thmref{Arb1}.
\begin{thm}  Suppose that $Z$ is the random field appearing in \eqref{newn} where the function $T\to h(T)$ is absolutely continuous and the function
$h'(T)$ is $L^2$ on each interval $[a,T_0]$, $a>0$.  Then for market price of risk $\lambda$ of the form
$$\lambda(t,T)=\int_0^T\eta(t,u, \omega )du;\hskip.4in\  T \leq T_0,$$
where $\eta$ is $\sFt$-predictable, $\sFt$ is a usual filtration with respect to which the white noise $\Wm$ in \eqref{newn} is measurable, and
$$\EP\exp\left(\int_0^{T_0}\int_0^{T_0}\big[\eta(t,u)\lambda(t,u)\int_u^{T_0}h'(\tau)^2d\tau/2+h(u)^2\eta(t,u)^2\big]\  dudt \right)<\infty,$$
the bond model \eqref{HalfSPDE} is free from arbitrage over the time interval $0\leq t\leq T_0$.
\label{Arb2}
\end{thm}

\lemref{STCOM} applies just as easily in this case (in fact see Theorem 2.2 in \cite{A} which applies to a much larger class of noises) and the proof is
quite similar to that for \thmref{Arb1} and will be omitted.

\end{document}